\documentclass[12pt,aps,nofootinbib,preprint,superscriptaddress]{revtex4}

\usepackage{amssymb}
\usepackage{amsmath}
\usepackage{amsfonts}

\newcommand{\eps}{\varepsilon}

\newcommand{\Ref}{Ref.}

\newcommand{\ie}{\emph{i.e.}}
\newcommand{\eg}{\emph{e.g.}}

\newcommand{\CP}{\emph{CP}}
\DeclareMathOperator{\diag}{diag}

\newcommand{\be}{\begin{equation}}
\newcommand{\ee}{\end{equation}}
\newcommand{\bea}{\begin{eqnarray}}
\newcommand{\eea}{\end{eqnarray}}

\begin{document}

\title{Neutrino oscillation parameter sampling with MonteCUBES}

\author{Mattias Blennow}
\email[]{blennow@mppmu.mpg.de}
\affiliation{Max-Planck-Institut f\"ur Physik (Werner-Heisenberg-Institut),
F\"ohringer Ring 6, 80805 M\"unchen, Germany}
\author{Enrique Fernandez-Martinez}
\email[]{enfmarti@mppmu.mpg.de}
\affiliation{Max-Planck-Institut f\"ur Physik (Werner-Heisenberg-Institut),
F\"ohringer Ring 6, 80805 M\"unchen, Germany}

\begin{abstract}
We present MonteCUBES (``Monte Carlo Utility Based Experiment
Simulator''), a software package designed to sample the neutrino
oscillation parameter space through Markov Chain Monte Carlo
algorithms. MonteCUBES makes use of the GLoBES software so that the
existing experiment definitions for GLoBES, describing long baseline
and reactor experiments, can be used with MonteCUBES. MonteCUBES
consists of two main parts: The first is a C library, written as a
plug-in for GLoBES, implementing the Markov Chain Monte Carlo algorithm
to sample the parameter space. The second part is a user-friendly
graphical Matlab interface to easily read, analyze, plot and export
the results of the parameter space sampling.
\end{abstract}


\preprint{MPP-2009-30}
\preprint{EURONU-WP6-09-03}

\maketitle

\section{Introduction}

After the firm establishment of neutrino oscillations from solar
\cite{Cleveland:1998nv,Abdurashitov:1999zd,Hampel:1998xg,Fukuda:2001nj,Ahmad:2001an,Ahmed:2003kj},
atmospheric \cite{Fukuda:1998mi,Ambrosio:2001je}, reactor
\cite{Apollonio:1999ae,Apollonio:2002gd,Boehm:2001ik,Eguchi:2002dm}
and accelerator \cite{Ahn:2002up,Aliu:2004sq,Michael:2006rx}
experiments, neutrino physics will enter a new precision age with
future facilities aiming to measure the subleading unknown mixing
parameters, such as $\theta_{13}$ and the \CP-violating phase $\delta$
\cite{Bandyopadhyay:2007kx}. These forthcoming neutrino oscillation
experiments, with unprecedented sensitivities, might also test new
physics in the neutrino sector beyond the present constraints. It is
then desirable to study the sensitivities of neutrino oscillation
experiments to these non-standard parameters together with the
improved constraints on the known mixing angles and mass squared differences
and the possible correlations between the two.

The task of studying the combined sensitivity of neutrino facilities
to both standard and new physics parameters and the degeneracies
between them becomes prohibitively expensive in computer time as the
number of parameters increases.  Thus, it is common to ``switch on''
these new physics parameters only one or two at a time, which does not
allow the exploration of correlations and degeneracies among them. In
order to address this problem, we propose the use of Markov Chain
Monte Carlo (MCMC) algorithms to explore these large parameter spaces
(see, \eg, \Ref~\cite{mcmcs}). Contrary to sampling the
$N$-dimensional parameter space through grids with $n$ samplings per
parameter, which require $\mathcal O(n^N)$ evaluations of the expected
number of events and $\chi^2$ for the considered setup, the number of
computations that MCMC sampling requires in order to achieve good
convergence grows at most polynomically with $N$.

MonteCUBES \cite{mcb} (``Monte Carlo Utility Based Experiment
Simulator'') contains a C library plug-in to implement MCMC sampling
into the GLoBES \cite{Huber:2004ka,Huber:2007ji} software. It thus
benefits from the flexibility of GLoBES in defining different
experiments while implementing an efficient scanning of large
parameter spaces. In addition to the MCMC sampling, MonteCUBES
includes an intuitive and user-friendly graphical Matlab interface to
interpret, plot and export the results of the MCMC sampling. It also
incorporates a method to locate degenerate solutions in a simple way
in order to tune the step proposal function of the MCMC sampling to
efficiently explore all the degenerate solutions, with the correct
weights, in the same run.

Not exclusive to the MCMC oriented functionalities of MonteCUBES, we
have added two extra utilities. The first is the implementation of two
new physics scenarios: the NonUnitarity Engine (NUE), which allows the
treatment of parameters describing the deviation from unitary of the
leptonic mixing matrix; and the non-Standard Interaction Event
Generator Engine (nSIEGE), which describes non-standard neutrino
interactions in matter. Calling these engines adds the dependence of
the oscillation probabilities on the extra parameters along with
useful functions in order to set their values and fix or free the
several new parameters in the minimization algorithms and the MCMC
samplings. The second functionality added is the possibility of
specifying the observed number of events of a given experiment instead
of computing them from the assumed ``true'' oscillation
parameters. This allows the usage of GLoBES and MonteCUBES to analyze,
not only forecasted data, but also real experimental data and study
the resulting constraints on the neutrino oscillation
parameters. Clearly, this feature requires extreme care when designing
the experiment definition files in order for the experiment definition
to coincide with the actual experiment.

\section{Markov Chain Monte Carlos}

Parameter determination through MCMC methods are based on Bayesian
inference. The aim is to determine the probability distribution
function (PDF) of the different model parameters $\theta$ given
some data set $d$, \ie, the \emph{posterior} probability
$P(\theta|d)$. From Bayes' theorem we have:
\begin{equation}
P(\theta|d)=\frac{P(d|\theta)P(\theta)}{P(d)} \equiv \frac{L_d(\theta) \pi(\theta)}{M}.
\end{equation}
The starting point is the \emph{likelihood} $L_d(\theta)=P(d|\theta)$,
\ie, the probability of observing the data set $d$ given certain
values of the parameters $\theta$. The \emph{prior}
$\pi(\theta)=P(\theta)$ is simply the probability of the parameters
having the value $\theta$ regardless of the data $d$, \ie, our
previous assumed knowledge of the parameters. Finally, the
\emph{marginal} probability $M$ is the probability $P(d)$ of
measuring the values $d$. It does not depend on the parameters
$\theta$ and can be disregarded as a normalization constant $M = \int
L_d(\theta) \pi(\theta) d \theta$, which cancels when comparing the
relative probabilities of different parameter values through the ratio
of the posteriors.\footnote{This normalization is,
  however, the key parameter in Bayesian model selection
  \cite{modelselection}.}

Thus, in order to compare the relative \emph{posterior} PDFs of
different sets of parameters $\theta_1$ and $\theta_2$ given the data
$d$, we must compute the ratio
\begin{equation}
\frac{L_d(\theta_1) \pi(\theta_1)}{L_d(\theta_2) \pi(\theta_2)}.
\end{equation}
The $\chi^2$ functions defined in GLoBES actually provide the
logarithm of the \emph{likelihood} of the data $d$ following a Poisson
distribution normalized to the distribution with mean $d$. Thus, computing
the exponential of the difference between these GLoBES functions for
$\theta_1$ and $\theta_2$ gives the desired \emph{likelihood}
ratio.

\subsection{The algorithm}

The aim of the MCMC is to create a Markov Chain that has the desired
distribution (the \emph{posterior} PDF for the oscillation parameters)
as its equilibrium distribution. The most popular implementation, and
the one used in MonteCUBES, is the Metropolis--Hastings algorithm. At
each step, the chain moves from a point in the parameter space
$\theta_1$ to another $\theta_2$ with a transition probability
$T(\theta_1,\theta_2)$. This transition probability is the product of
the proposal function $W(\theta_1,\theta_2)$ times the probability of
accepting the new step:
\begin{equation}
\alpha(\theta_1,\theta_2) = \min\left(1,\frac{P(\theta_2|d)W(\theta_2,\theta_1)}{P(\theta_1|d)W(\theta_1,\theta_2)}\right).
\end{equation}
This algorithm ensures that detailed balance 
\begin{equation}
P(\theta_2|d)T(\theta_2,\theta_1)=P(\theta_1|d)T(\theta_1,\theta_2)
\end{equation}
is satisfied (while maximizing the acceptance) and thus, $P(\theta|d)$
is the equilibrium distribution of the chain.

In the original Metropolis algorithm, as well as in the default
implementation of MonteCUBES, the proposal function
$W(\theta_1,\theta_2)$ is symmetric so that the probability $\alpha(\theta_1,\theta_2)$ of
accepting the new proposed step is simply given by the ratio of the
\emph{posteriors}, \ie, the exponential of the difference of the
$\chi^2$ provided by GLoBES plus the \emph{prior}.

In addition to using Gaussian proposal functions, MonteCUBES provides
an easy way of treating degeneracies by changing the proposal function
by randomly adding or subtracting steps with the correct length in the
direction between the degeneracies. MonteCUBES can automatically
search for degeneracies by increasing the temperature of the chain $T$
so that the likelihood is modified to $L_T \propto
L^{1/T}$. This procedure flattens the likelihood distribution
so that the chains can move between the different degeneracies. The
temperature and step sizes are then decreased in successive steps and
thus the different chains get stuck around different minima, unable to
move through the disfavored regions when $T$ is too low. Finally, the
points where the different chains have stopped are compared to decide
how many different minima the chains have fallen into and a
minimization of the log-likelihood is performed from those starting
values so that the minima are reached. The difference between the
minima, located in this manner, can then be used in the proposal function.
Thus, when performing the MCMC sampling of the parameter space, the
chains can jump freely between the degeneracies and sample them with
the correct relative weights. Finally, the standard MonteCUBES
proposal function can be replaced by an arbitrary user-defined
proposal function, which does not necessarily need to be symmetric (as
long as the user also implements the proper transition ratio
function).

\subsection{Interpreting the results}

To test the convergence of the chains we use the method proposed in
\Ref~\cite{conv} with convergence criteria that can be specified by
the user. The key parameter that controls the the convergence speed
and how well the chains will sample the distribution is the typical
step sizes of the proposal function $W(\theta_1,\theta_2)$. Optimal
step sizes are of the order of the expected $1 \sigma$ allowed region.

If the steps are too small, the chains will sample small areas of the
parameter space and take a very long time to cover the whole region of
interest. Different chains will sample different regions, depending on
their starting values, and will give different estimates of the means
of the parameters, translating to very bad convergence. A good
diagnose of too short a step in a given parameter is a long
correlation length between the values sampled for this parameter as
the chains progresses.

If the steps are too large, the chains will often propose jumps to
regions of the parameter space which are very disfavored and the
probability of accepting the step will be very small, requiring a long
time to accumulate enough statistics to properly analyse the sampled
probability. A good diagnose of too large steps is that the chains
stop too many times at each accepted point.

The output of the MCMC sampling is several chain files, containing a
list of the accepted points in the parameter space together with the
weight (the number of times the chain stayed at that point), since the
equilibrium density of the chain is $P(\theta)$ the density of points
in the parameter space given by the chains will be proportional to the
\emph{posterior} probability we want to sample. Thus, simply binning
the parameter space and distributing the points of the chains in their
corresponding bins will provide the relative \emph{posterior}
probabilities of each bin in the multidimensional parameter
space. Notice that the marginalization over nuisance parameters does
not require a time-consuming minimization but is simply achieved by
ignoring the corresponding parameters, effectively projecting the
\emph{posterior} PDF to the parameter subspace of interest.

\section{The graphic user interface}

Even if the interpretation and analysis of the chains is
straightforward, as outlined in the previous section, processing them
can be cumbersome. For this task MonteCUBES includes an intuitive
Matlab Graphic User Interface (GUI) that allows to read, combine,
analyze, plot and export the results of the MCMC sampling.  After
opening the GUI the user can select the results of the run to be
analyzed by opening the corresponding summary file generated by
MonteCUBES together with the chains.  The summary file contains the
relevant information on the number of chains, number of samples, free
parameters and convergence criteria required to properly read and
analyze the results of the chains. After reading the summary file the
GUI reads and processes the corresponding chains. The user can then
either combine chains from further runs to increase the statistics of
the chains or plot the results in several ways. The following plotting
options are included in the GUI:
\begin{itemize}

\item{\bf 1D Histogram:} This plots a histogram of the number of
  points in the chains as a function of the parameter selected by the
  user in a range and with a number of bins that can be
  user-specified. The main application of this plot is to diagnose how
  well the chains have converged and sampled the chosen parameter.
  The histogram should resemble a Gaussian centered at the most likely
  value of the parameter or some multimodal distribution if that
  particular parameter presents degeneracies.

\item{\bf 1D chain progression:} This plot also diagnoses the
  convergence and how well a parameter has been sampled. It plots the
  consecutive values of a parameter which the chains have visited.
  For well sampled and converged chains these plots should resemble
  ``white noise'', any residual correlation length in the form of
  oscillations in these plots indicate a poor sampling of the
  parameter and the typical size of the steps in the proposal function
  may need to be changed (typically increased) to achieve better
  sampling. Notice that the chain convergence can be defined in a
  quantitative way and that MonteCUBES can test the convergence of all
  the parameters in the chains and continue the sampling until 
  the desired level of convergence is achieved. However,
  these plots constitute a useful tool to determine how to tune the steps in
  the proposal function to speed up the convergence of the chains and
  obtain better sampling of the parameter space.

\item{\bf 1D confidence region:} This plots the posterior
  distribution marginalized in all the parameters except the one
  selected and highlights the most favored region at a user-defined
  confidence level.

\item{\bf 2D scatter:} This draws a scatter plot in the specified
  parameters of the points which the chains have sampled. This can
  also be a good diagnosis of bad sampling, since the plot should be
  uniform with clearly visible denser regions corresponding to the
  best fit values that gradually thin when moving away from the
  favored area.

\item{\bf 2D confidence contours:} This plots the isoprobability
  contours at the specified confidence levels for the specified
  parameters.

\item{\bf Triangle plot:} This option plots together the 1D confidence
  region and all the possible 2D confidence contours for a set of
  selected parameters.

\item{\bf 3D surface:} This plots the isoprobability surface in the
  specified three parameters at the chosen confidence level.

\end{itemize}
For all of the above plotting options, except the two
latter\footnote{All the plots from the triangle plot can be reproduced
  using the 1D confidence region and 2D confidence contour plots,
  while the 3D surface plot is not really suitable for the exporting
  of data.}, there is also the option of exporting the high-level data
for the plots into text files. This functionality is provided so that
the user can plot the data in a different graphical program if
desirable.

In addition, all of the plotting functions provide the user with the
choice of plotting the results against arbitrary transformations of
the oscillation parameters. Since the Jacobian of these
transformations may not be constant, the user-interface also provides
the possibility of specifying an arbitrary weight function in order to
compensate. This also allows to change the prior post-simulation.

\section{New physics implementations}

The MonteCUBES distribution includes two GLoBES implementations of new
physics. Both of these contain nine extra parameters in addition to
the six standard neutrino oscillation parameters. Thus, these
scenarios are ideally suited for MCMC exploration of the parameter
space, since a full scan quickly becomes inefficient.

The first implementation is the NonUnitarity Engine (NUE), which can
be used together with GLoBES and MonteCUBES functions in order to
include deviations from unitarity of the lepton mixing matrix
parameterized in a completely general way. A non-unitary lepton mixing
matrix in the charged-current interaction between neutrinos and
charged leptons is a generic feature of models involving extra degrees
of freedom that can mix with either of the left-handed lepton components
\cite{Langacker:1988ur}. In particular, in the popular type-I seesaw
models that accommodate the smallness of neutrino masses through the
addition of heavy fermion singlets (right-handed neutrinos), these
extra degrees of freedom will mix with the light active neutrinos,
giving rise to a larger mixing matrix than the standard $3\times 3$
matrix.  The $3 \times 3$ submatrix describing the mixing among the
light mass eigenstates, accessible at low energies, and the three
active flavour eigenstates will, in general, not be unitary. In
standard seesaw models, this unitarity violation is expected to be
unobservably small. However, these violations are induced by
a lepton number conserving operator independent of the one that
generates neutrino masses.  The smallness of the neutrino mass can
then be naturally accommodated through a slightly broken lepton number
symmetry, as in the inverse or double seesaw models, with large
potentially testable deviations from unitarity of the lepton mixing
matrix.

A convenient way of parameterizing the effects of a non-unitary mixing
in neutrino oscillations is splitting the general non-unitary matrix
$N$ as the product of an Hermitian times a unitary matrix
\cite{FernandezMartinez:2007ms} $N = (1+\eps)U$, where
$\eps^\dagger = \eps$. Since strong constraints can be derived
on the unitarity deviations through electroweak decays
\cite{Nardi:1994iv,Tommasini:1995ii,Antusch:2006vwa,Antusch:2008tz},
$\eps$ should be a small perturbation and $U \simeq U_{\rm PMNS}$. The
NUE adopts this parameterization, adding the six extra moduli and
three extra phases included in the Hermitian $\eps$ to the
standard parameters of the unitary part of the general mixing matrix.

By using NUE, the oscillation probabilities are modified so that the
dependence on these extra nine parameters is taken into account. The
engine also includes useful functions in order to set the values of
these parameters in the parameter vectors and fix or free them in
GLoBES' minimization functions or in MonteCUBES' MCMC sampling.

The second new physics implementation is the non-Standard Interaction
Event Generator Engine (nSIEGE). This engine is designed to treat
non-standard neutrino interactions (NSI) with matter in their most
general form. The formalism of NSI parametrises the effects from
physics beyond the Standard Model on neutrino interactions through
effective four-fermion operators
\begin{equation}
 \mathcal L_{\rm NSI} = -2\sqrt{2}G_F \eps^{fP}_{\alpha\beta}
               [\bar f \gamma^\mu P f][\bar \nu_\alpha \gamma_\mu P_L \nu_\beta],
\end{equation}
where $f$ is a matter fermion, $P$ is either a left- or right-handed
projector, and $\eps_{\alpha\beta}^{fP}$ parameterizes the strength of
the NSI relative to the standard weak interactions. The new
interactions give rise to non-standard matter interaction terms in the
neutrino oscillation formalism, effectively leading to the replacement
\begin{equation}
 H_{\rm matter} = \sqrt 2 G_F N_e \diag(1,0,0)
  \to \sqrt 2 G_F N_e [\diag(1,0,0) + \eps]
\end{equation}
of the matter interaction term in the neutrino oscillation
Hamiltonian. Here, $\eps$ is a Hermitian matrix and $N_e$ is the
electron number density.

This type of new physics has already been studied extensively with
GLoBES
\cite{Blennow:2007pu,Kopp:2007mi,Kopp:2007ne,EstebanPretel:2008qi,Kopp:2008ds,Blennow:2008ym,Blennow:2008eb,Winter:2008eg,Malinsky:2008qn}
by several authors, but usually by fixing most of the parameters. The
nSIEGE implementation is very similar to the that of the NUE in terms
of the API.

\section{Conclusions}

We have presented a new software tool, MonteCUBES, which allows the
exploration of the neutrino oscillation parameter space through Markov
Chain Monte Carlo sampling. The MCMC algorithms are far more efficient
than minimizations or grids over large parameters spaces and we
therefore believe MonteCUBES to be a particularly powerful tool for
the investigation of the effects of new physics in neutrino
oscillations, since they imply the addition of new parameters to the
already high-dimensional standard parameter space. The simulation part
of MonteCUBES is designed as a plug-in for the GLoBES software and
thus benefits from a very flexible experiment definition. It includes
a useful method to find degeneracies and allow a faster scan, taking
all the allowed regions detected into account in the same sampling. We
have also developed an intuitive graphic user interface for Matlab,
which allows to easily read, combine, analyze, plot and export the
results of the MCMC exploration of the parameter space in order to
interpret the constraints that a given experiment can derive on the
oscillation parameters.

Apart from the MCMC sampling, we incorporate two useful
functionalities in the MonteCUBES software, both of them are
compatible with the new MonteCUBES functions and the original
functions defined in GLoBES. The first one is the implementation of
two scenarios of new physics, the NonUnitarity Engine (NUE) that
incorporates all the extra parameters required to study the effects
that a deviation from unitarity of the lepton mixing matrix will have
in neutrino oscillations, and the non-Standard Interaction Event
Generator Engine (nSIEGE) that describes non-standard neutrino
interactions in matter. These engines are examples of the kind of
applications we believe MonteCUBES to be best suited to, namely the
exploration of large parameter spaces. In both cases, six extra real
parameters and three extra phases are required in order to take the
new physics into account in the most general setting. Added to the six
standard neutrino oscillation parameters, this results in
15-dimensional parameter spaces.
Using MonteCUBES, a full scan of the unitarity violation parameter
space in order to test the sensitivity of a Neutrino Factory~\cite{Geer:1997iz,DeRujula:1998hd} to this
particular scenario of new physics has been performed in
\Ref~\cite{NUpaper}. 

The second additional functionality is the possibility to input the
data of a given experiment so that GLoBES and MonteCUBES can be used
to analyze real data and not only forecast the sensitivities by
specifying some ``true'' oscillation parameters and predicting the
event rates to be observed from them.

We conclude that the MCMC methods implemented in MonteCUBES constitute
a powerful tool to explore the bounds on the neutrino oscillation
parameters that different experimental setups can give, as well as
possible degeneracies and correlations among them. In particular, they
allow an easy and efficient way of treating all neutrino oscillation
parameters simultaneously, as well as including additional parameters
from different non-standard physics.

\begin{acknowledgments}

We would like to thank M.~Beltran, P.~Coloma, A.~Donini, J.~Kopp and L.~Verde for
very useful discussions. This work was supported by the Swedish Research Council
(Vetenskapsr\aa{}det), contract no.~623-2007-8066 [M.B.]. E.F.M.~acknowledges support
from the European Community under the European Commission Framework Programme 7 Design Study:
{EUROnu}, Project Number 212372. The authors also acknowledge support by the DFG cluster of excellence ``Origin
and Structure of the Universe''.

\end{acknowledgments}

\end{document}